\documentclass[fleqn,usenatbib]{mnras}
\usepackage[T1]{fontenc}
\usepackage{verbatim}
\usepackage{newtxtext,newtxmath}

\DeclareRobustCommand{\VAN}[3]{#2}
\let\VANthebibliography\thebibliography
\def\thebibliography{\DeclareRobustCommand{\VAN}[3]{##3}\VANthebibliography}

\usepackage{graphicx}
\usepackage{amsmath}
  
\usepackage{bm}
\usepackage{subfig}
\usepackage{afterpage}
\usepackage{mathtools} 
\usepackage{float}
\usepackage{times}
\usepackage{color}
\usepackage{amsfonts}
\usepackage{booktabs}
\usepackage{siunitx}
\usepackage{soul}
\usepackage{xcolor}
\usepackage{float}
\usepackage{capt-of}
\usepackage{multicol}
\usepackage{multirow,multicol}
\usepackage{placeins}
\usepackage[normalem]{ulem}

\newcommand{\ie}{i.e.,~}

\title{Nonthermal hotspot orbiting the supermassive black hole  Sgr\,A*}

\author[Cruz-Osorio et. al.]
{Alejandro Cruz-Osorio$^{1}$\thanks{E-mail: aosorio@astro.unam.mx},
David Barrero-González$^{1}$,
Juan J. Zaldivar$^{2}$,
Yosuke Mizuno$^{3,4}$\\
$^{1}$Instituto de Astronom\'{\i}a, Universidad Nacional Aut\'onoma de M\'exico, AP 70-264, Ciudad de M\'exico 04510, M\'exico \\
$^{2}$Instituto Nacional de Astrofísica, Óptica y Electrónica, Santa María Tonantzintla, Puebla, 72840, Mexico\\
$^{3}$Tsung-Dao Lee Institute, Shanghai Jiao Tong University, Shanghai, 201210, People's Republic of China\\
$^{4}$School of Physics and Astronomy, Shanghai Jiao Tong University, Shanghai, 200240, People’s Republic of China
}

\usepackage{hyperref}
\hypersetup{
  colorlinks=true,        % false: boxed links; true: colored links
  linkcolor=blue,         % color of internal links
  citecolor=magenta,         % color of links to bibliography
  filecolor=magenta,      % color of file links
  urlcolor=magenta        % color of external links
}

\begin{document}
\label{firstpage}
\pagerange{\pageref{firstpage}--\pageref{lastpage}}
\maketitle

\begin{abstract}
   Recent millimeter wavelength observations at 230 GHz by the Event Horizon Telescope 
    have resolved the innermost structure of the Galactic Center, Sgr\,A*, revealing a ring-like 
    structure consistent with synchrotron emission from a magnetized torus surrounding a 
    supermassive black hole. These observations reveal two key features: three prominent 
    hotspots near the photon ring and low light-curve variability, with a modulation index of  
    $\bar{\sigma}/\bar{\mu} \lesssim 0.1$.
    State-of-the-art general relativistic magnetohydrodynamic and radiative transfer simulations 
    cannot simultaneously reproduce the three hotspots and low light-curve variability observed 
    in Sgr\,A*. We propose that an orbiting hot blob, formed in the accretion disk, 
    could account for hotspots. We systematically explore its physical properties, 
    including size, emission amplitude, orbital radius, eccentricity, velocity, and orbital 
    plane orientation.
    The results -- applying two observational constraints--show that the preferred model 
    corresponds to a black hole with dimensionless spin $a_{\star}=0.5$, a magnetically 
    arrested disk configuration for the magnetic field, and the presence of a hotspot with 
    size $\sigma_{\rm hs}=2\,M$ and intensity amplitude $A=10$, orbiting circularly at 
    $r_{\rm orb}=6\,M$ with Keplerian velocity in the black hole’s equatorial plane, and 
    an  inclination view angle of $i^\circ=20$. 
\end{abstract}


%%%%%%%%%%%%%%%%%%%%%%%%%%%%%%%%%%%%%%
%%%%%%%%%%%%%%%%%%%%%%%%%%%%%%%%%%%%%%
\section{Introduction} 
\label{sec:intro}

    At the heart of our Galaxy lies Sagittarius A* (Sgr\,A*), a compact radio source 
    identified as a supermassive black hole (SMBH) with a mass of approximately 
    $M=4.14 \times 10^6M_{\odot}$, located at a distance of $D\sim 8.127 kpc$ 
    \citep{Gravity2019, EHT_SgrA_PaperI}. Multi-wavelength observations have 
    long hinted at the complex dynamics of the innermost regions of the accretion 
    flow \citep{Lu2011,Michail2021,EHT_SgrA_PaperII,Michail2024}. However, it 
    is the recent breakthrough achieved by the Event Horizon Telescope (EHT) 
    that has provided an unprecedented view of this region. The EHT observations 
    at $230 {\rm GHz}$ revealed a bright, asymmetric ring structure encircling a 
    central shadow -- an imprint of the black hole's strong gravity \citep{EHT_SgrA_PaperI}.\\
    Alongside this striking morphology, significant variability has been detected on 
    timescales ranging from tens of minutes to a few hours \citep{EHT_SgrA_PaperIV}, 
    suggesting that the plasma surrounding the black hole is highly dynamic and 
    evolves on rapid timescales.\\
    One proposed explanation for the observed variability is the presence of compact, 
    bright regions --commonly referred to as “hotspots”-- orbiting near the 
    event horizon \citep{Matsumoto2020,Nathanail2020,Porth2021,Dexter2020b,Yfantis2024I}. These 
    structures may naturally arise from magnetic reconnection events within the 
    turbulent, magnetized plasma, or from instabilities in the innermost regions of 
    accretion flow \citep{deGouveia_2010,RodriguzRamirez_2019}. If their orbital 
    motion can be detected, it could provide valuable constraints on the  spacetime 
    geometry and the spin of the black hole, as well as a direct probe of plasma 
    physics in the strong-field regime \citep{Ozel_2022}.\\
    
    Theoretical studies have demonstrated that the appearance and variability of 
    such hotspots are highly sensitive to their intrinsic properties (e.g., size, emissivity), 
    their orbital parameters (e.g., radius, inclination, eccentricity), as well as global 
    factors such as the black hole’s spin and the observer’s viewing angle.
    A key challenge in modeling these phenomena lies in the accurate treatment 
    of nonthermal electron populations, which are expected to dominate the 
    high-frequency synchrotron emission. Recent advances in general relativistic 
    magnetohydrodynamic (GRMHD) and general relativistic radiative transfer 
    (GRRT) modeling have enabled the self-consistent exploration of such scenarios 
    by coupling fluid dynamics with radiative transfer, incorporating physically 
    motivated prescriptions for the electron distribution function. However, current 
    GRMHD and GRRT simulations still face difficulties in simultaneously 
    reproducing both the observed ring-like morphology and the low variability 
    of Sgr\,A* suggesting that additional localized emission processes—such as 
    orbiting nonthermal hotspots—may play a critical role 
    \cite{Ripperda2019b,Nathanail2020,Ripperda2020,Nathanail2021b,Scepi2021,
    Ripperda2022,Vos2024b}.\\
    
    In this work, we use GRMHD simulations performed with the BHAC code 
    ~\citep{Porth2017,Olivares2019} and compute synthetic synchrotron images 
    and light curves using the GRRT code BHOSS ~\cite{Younsi2012}. We 
    model a nonthermal hotspot orbiting the supermassive black 
    hole at the Galactic Center and systematically explore the influence of various 
    physical and geometric parameters on the observed variability at 230 GHz. 
    In particular, we investigate the effects of orbital radius, eccentricity, inclination, 
    hotspot size and emissivity, the black hole’s spin, and the observer’s viewing 
    angle. Our results show that the modulation index—a proxy for flux variability—
    depends strongly on the spin and viewing geometry, with important implications 
    for the detectability of such features by current and future VLBI observations.
    
    The structure of the paper is as follows: in Section \ref{sec:simulations}, we
    describe the numerical setup of our  GRRT calculations and describe the 
    toy model of orbiting hotspot around the rotating black hole. Section 
    \ref{sec:results} present the results
    of our parameter study, highlighting the regimes that maximize variability. 
    Finally, in Section \ref{sec:conclusions}, we summarize our main findings and
    discuss their implications for future time-resolved observations of Sgr\,A*.
    
    \section{Numerical modeling of the black hole shadow and orbiting hotspot} 
    \label{sec:simulations}
    
    The synthetic image of the galactic center, Sgr\,A*, presented 
    in this paper was modelled by the synchrotron emission from 
    evolved black hole magnetized accretion disk system. The 
    astrophysical plasma evolution is studied with a state-of-the-art 
    general relativistic magnetohydrodynamics code, 
    \texttt{BHAC} ~\citep{Porth2017,Olivares2019}. The 
    shadow and electromagnetic emission from the gas are 
    generated with the general relativistic radiative transfer 
    \texttt{BHOSS} code ~\cite{Younsi2012}.  
    The black hole spacetime is assumed to be described by 
    Kerr solution in general relativity, here we explore the cases 
    for dimensionless spin parameter $a_{\star}= -0.94,\, -0.50,\, 
    0.00,\, 0.50\,$ and $0.94$. The plasma in the accretion disk 
    at the initial time is constructed by Fishbone-Moncrief solution, 
    which assumes a perfect fluid in gravito-hydrostatic equilibrium 
    with constant specific angular momentum $\ell:=u_\phi/u_t = 6.76$, 
    where $\boldsymbol{u}$ is the fluid four-velocity. 
    The gas in the accretion disk is described with an ideal equation of 
    state~\citep[see e.g.][]{Rezzolla_book:2013}, 
    with adiabatic index $\Gamma=4/3$. 
    The size of the torus is defined by its inner edge and central radius 
    $r_{\rm in} = 20\,M$ and $r_{\rm c}=40\,M$, while the external radii 
    are determined by gravitational potential ~ 
    \citep{Fishbone76,Font02b,Shiokawa2012,Rezzolla_book:2013}.  
    In order to keep the equilibrium, we introduce a weak 
    single-loop poloidal magnetic field, which is added in terms of the 
    vector potential $A_{\phi}\propto\mathrm{max} \{(\rho/\rho_{\rm max})
    \left(r/r_{\rm in}\right)^3\sin^3\theta\exp\left(-r/400\right)-0.2,0\}$ 
    and normalised to the minimum value of plasma beta, so that 
    $\beta_{\mathrm{min}}=(2p/b^2)_{\mathrm{min}}=100$ inside the disk.\\
    
    The synchrotron emission is computed solving general relativistic 
    radiative-transfer equation along null geodesics,
    where the electromagnetic radiation propagates ~
    \cite{Younsi2012}. The optical depth, magnetic fields, 
    temperature, and emission properties of the plasma are 
    obtained from our GRMHD simulations covering $5,000\ M$ 
    ($25,000\ M-30,000\ M$) which corresponds to $\sim 30\,  
    {\rm hours}$ of observations data, given the estimated mass 
    of Sgr\,A* of $4.14\times10^6\,\mathrm{M}_{\odot}$ and 
    source distance of $8.127\,{\rm kpc}$~\citep{EHT_SgrA_PaperI, 
    EHT_SgrA_PaperV}. The synthetic emission is normalized to match 
    the observed flux density of SgrA* at $230\,{\rm GHz}$, approximately 
    $2.4\,{\rm Jy}$, as reported by the EHT Collaboration 
    \citep{EHT_SgrA_PaperI}. The GRMHD simulations where 
    performed using the numerical domain of $r\in[1.18\,r_{\rm EH}
    ,\, 2500\,M]$, $\theta\in[0,\pi]$, and $\phi\in[0,2\pi]$, in particular 
    we use the logarithmic Kerr-Schild coordinates \cite{Cruz2020}.
    
    The estimation of the optical properties 
    require the knowledge of the electron temperature as well 
    as the electron distribution function. We use non-thermal 
    electron population function with power-law motivated in
    microphysics of the plasma simulations, \ie
    $\mathrm{d} n_{\rm e} /\mathrm{d} \gamma_{\rm e}
    = N \gamma_{\rm e} \sqrt{ \gamma_{\rm e}^{2}-1} \left[ 1 + (\gamma_{\rm
    e}-1)/(\kappa w) \right]^{-(\kappa+1)}$, where $\gamma_{\rm e}$  is the
    Lorentz factor of the electrons and $N$ is a normalisation parameter, 
    which naturally converge to the thermal emission for low magnetized plasma 
    \citep{Cruz2022,Roeder2023,Zhang2024}. 
    The weighted temperature is defined as $w:=(\kappa-3)\Theta_{\rm
    e}/\kappa+\tfrac{1}{2}\epsilon\left[1+\tanh{(r-r_{\rm inj})}\right]
    (\kappa-3)m_{\rm p}\sigma/(6 \kappa\ m_{\rm e})$, here $r_{\rm
    inj}=10\,M$ is the injection position, $\Theta_{e} ={p m_{\rm
    p}\mathcal{T}}/{ m_{\rm e} \rho}$ is the dimensionless electron
    temperature, $m_{\rm p}$ and $m_{\rm e}$ are respectively
    the proton and electron masses, where the ion-to-electron temperature 
    ratio is obtained from self-consistent particle-in-cell (PIC) simulations, 
    $\mathcal{T}(\beta, \sigma) 
    = (t_0 + t_1\, \sigma^{\tau_1} \tanh\left[t_2 \, \beta \,
    \sigma^{\tau_2}\right] + t_2\, \sigma^{\tau_3} \tanh\left[t_3 \, 
    \beta^{\tau_4}\, \sigma \right])^{-1}$, where $t_i = (0.4,\, 0.25,\, 
    5.75,\, 0.037)$, and $\tau_i=(-0.5,\, 0.95,\, -0.3,\, -0.05)$, 
    $\sigma=b^2/\rho$ is the magnetisation parameter, i.e. the ratio of the 
    magnetic to rest-mass energy densities, $\beta = 2p/b^2$ is the ratio of 
    the magnetic and gas pressure
    and $\epsilon=0.5$ is the fraction 
    of magnetic energy contributing to the heating of the radiating electrons. 
    The power-law index of the electron distribution function $\kappa$, 
    depends on the microphysics of the plasma so that
    $\kappa:=2.8+0.2\,\sigma^{-1/2}+1.6\,\sigma^{-19/100}\tanh{(2.25\,
    \sigma^{26/100} \beta)}$. 
    Both the ion-to-electron temperature ratio and the power-law index were 
    computed from turbulent PIC simulations presented in \cite{Meringolo2023,Imbrogno2024}. 
    In \cite{Meringolo2023}, the PIC simulations are initialized with a plasma in a 
    strong turbulent regime, characterized by $\delta B/B_{0} \sim 1$, within a 
    numerical domain of size $L = 2730\,d_{e}$, where $d_{e}$ denotes the electron 
    skin depth. Initially, electrons and ions are assumed to be in thermal equilibrium 
    and are described by Maxwell--Jüttner distribution function. The fitting formulas 
    adopted here were derived after two Alfvén crossing times when the system 
    reaches a quasi-steady state. For a similar procedure 
    applied to a single current sheet, see \cite{Ball2018a}.\\
    Following the same procedure as in \cite{Davelaar2018, Fromm2021b, Cruz2026}, 
    and as is widely adopted in the literature, we compute the emissivity and absorptivity 
    coefficients, $j$ and $\alpha$, as functions of the electron temperature and electron 
    distribution function, following the prescriptions of \cite{Pandya2016}. 
    Here, we assume that the eDF, as well as the emissivity and absorptivity
     coefficients, can be computed from the local magnetization, $\sigma$, and plasma $\beta$ measured 
     in the GRMHD simulations. However, this eDF may not provide an appropriate 
     description of the hotspot, since the hotspot is introduce ad hoc. Thus, during the ray-tracing 
    calculation, the local plasma properties 
    are evaluated at each point along the null geodesics, and the radiative transfer equations 
    are solved assuming that the fitting formulas are instantaneously applicable at each location.
    
    For comparison with observed flux of Sgr\,A* at 230\,GHz of 
    $\simeq 2.4\,{\rm Jy}$~\citep{EHT_SgrA_PaperI,EHT_SgrA_PaperV}, 
    we normalise the emission at a resolution of $400\times400$ pixels, 
    which corresponds to a field of view of $200\, \mu {\rm as}$.

    \begin{figure}
      \centering
      	\includegraphics[width=0.5\textwidth]{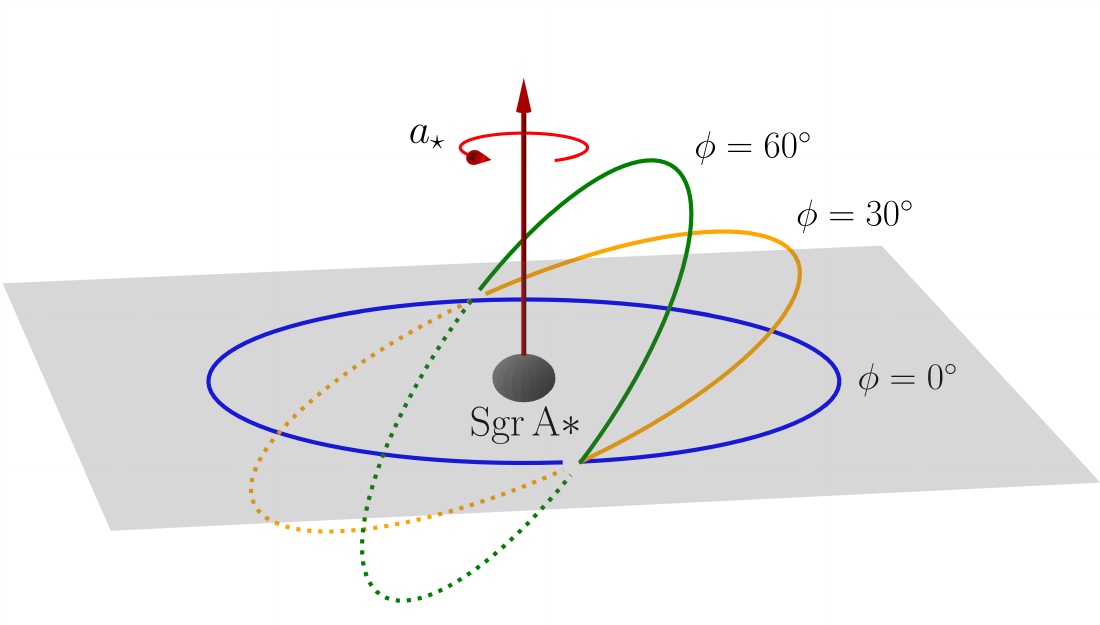}
     \caption{Toy model hotspot orbits around the galactic center, Sgr\,A*. 
     Here, $\phi$ represents the inclination angle of the hotspot orbital 
     plane relative to the black hole equatorial plane (grey), and $a_{\star}$ 
     is the black hole spin (red arrow). Blue, orange and green lines 
     are different orbits of the hotspot.
         \label{fig:orbit}}
    \end{figure}

    \subsection{Toy model of orbiting hotspot around Sgr\,A*}
    \label{sec:hotspot}
    
    \begin{figure*}
      \centering
      	\includegraphics[width=0.725\textwidth]{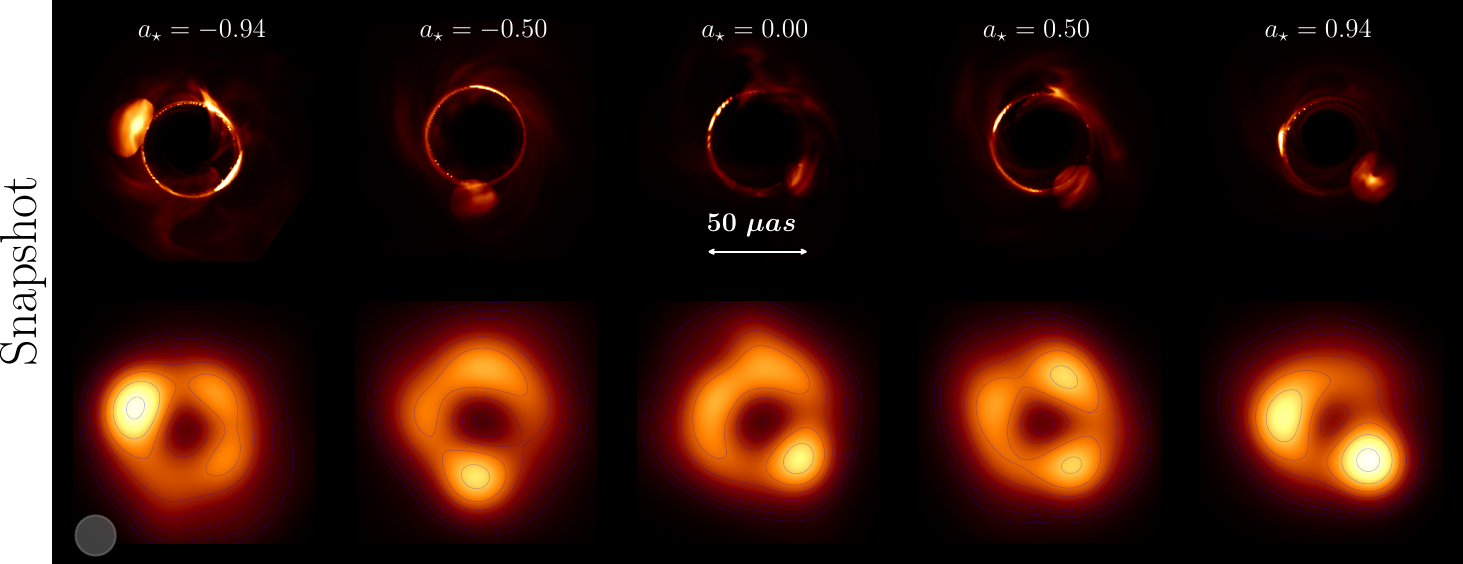}\\
      	\includegraphics[width=0.725\textwidth]{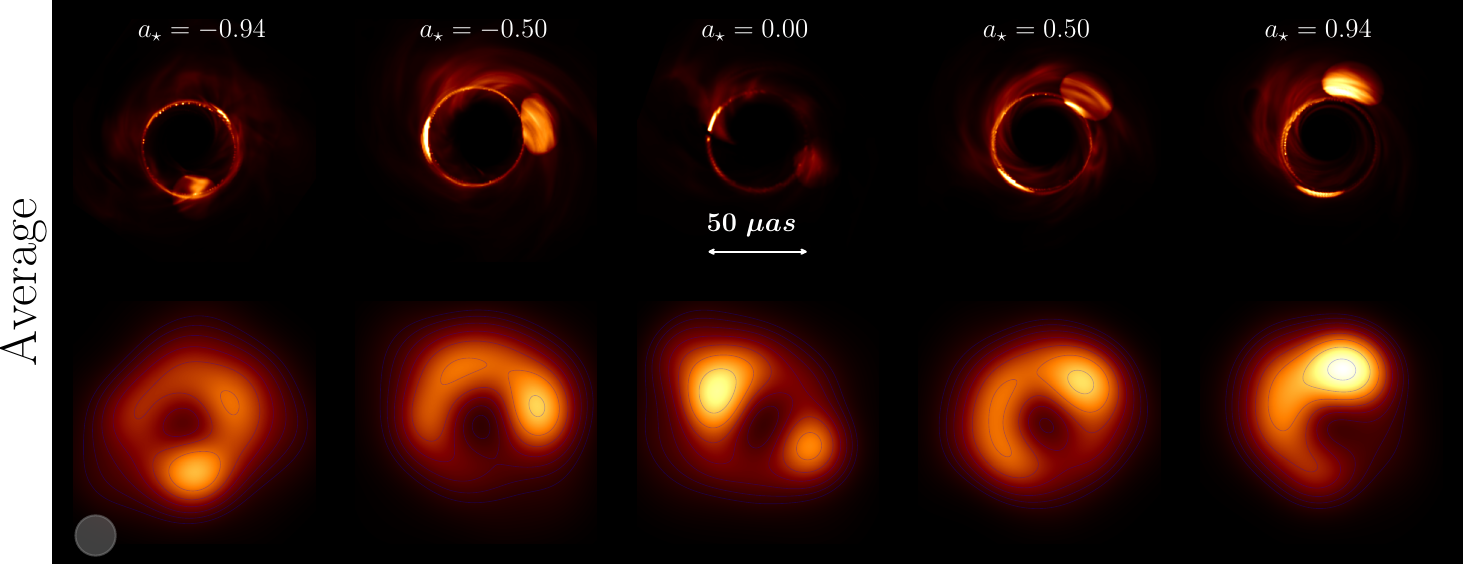}\\
     \caption{Synthetic black hole shadow images for a set of representative 
     models. From left to right, we show five black holes with dimensionless 
     spins, $a_{\star}= -0.94,\, -0.50,\, 0.00,\, 0.50\,$, and $0.94$, respectively. 
     The first row presents {\it snapshot} images, while the second row shows 
     the {\it time-averaged} images over a three-hour window centered on the 
     corresponding snapshot. The top panels display GRRT images, while 
     bottom panels show the corresponding convolved images with a Gaussian beam 
     matching the EHT angular resolution of $20\, \mu{as}$. 
     The inclination angle is $i=20^{\circ}$ for the black 
     hole with spin $a_{\star}= 0.50$ and $i=30^{\circ}$ for the other spins, \ie $a_{\star}\neq 0.50$. The 
     hotspot is assumed to orbit with Keplerian velocity at radius  
     $r_{\rm orb}=6.0$ and eccentricity $e=0.0$. The emission has an 
     amplitude $A=10$ and the size is $\sigma_{\rm hs}=2.0$. The 
     contours in the flux density correspond to 
     $S_{i}=S_{\rm min}+0.43\,{\rm mJy}\times\sqrt{2}^{i}$, where 
     $i=0,1,\ldots,n$, such that $S_{n}<S_{\rm max}$.
         \label{fig:rep}}
    \end{figure*}

    \begin{figure}
      \centering
      	\includegraphics[width=0.339\textwidth]{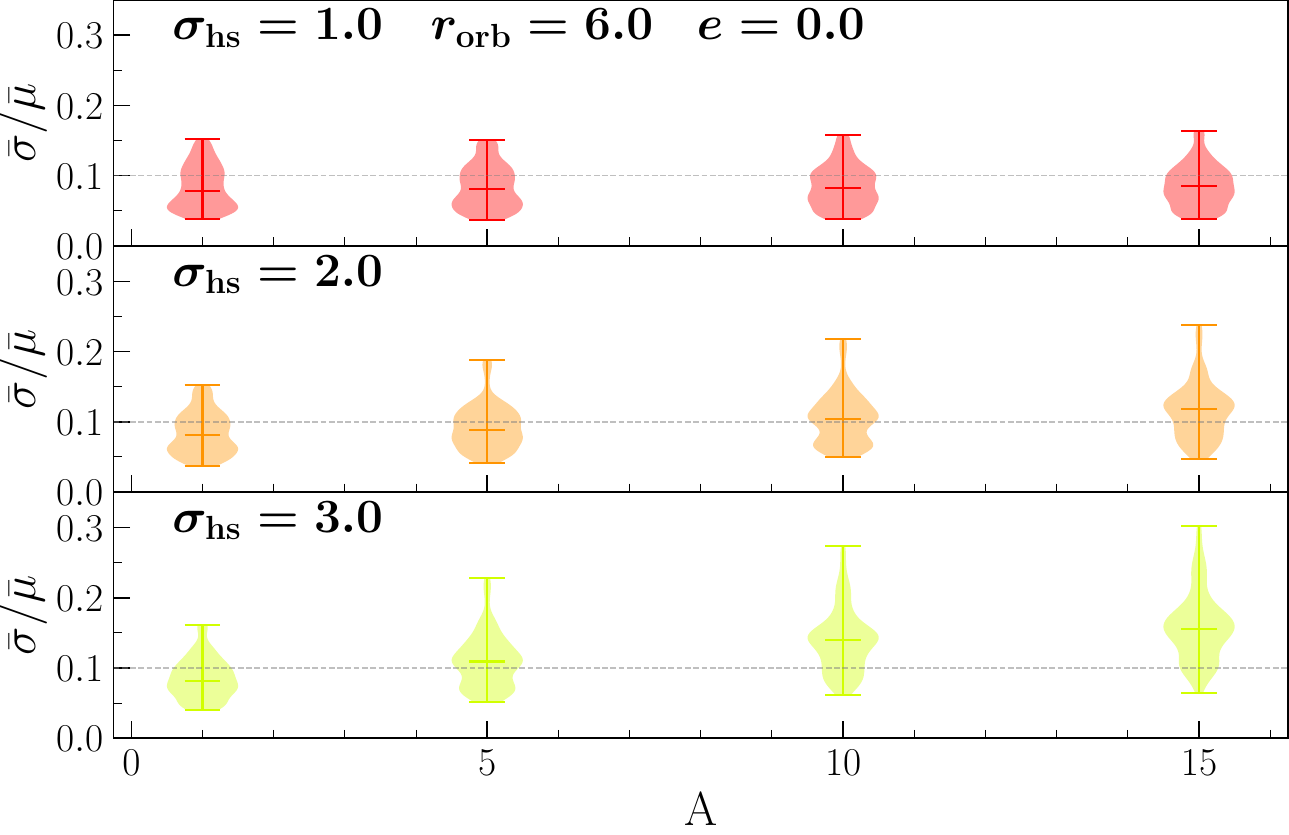} \begin{picture}(5,5) \put(0.5,60){\Large (a)} \end{picture}\\
      	\includegraphics[width=0.339\textwidth]{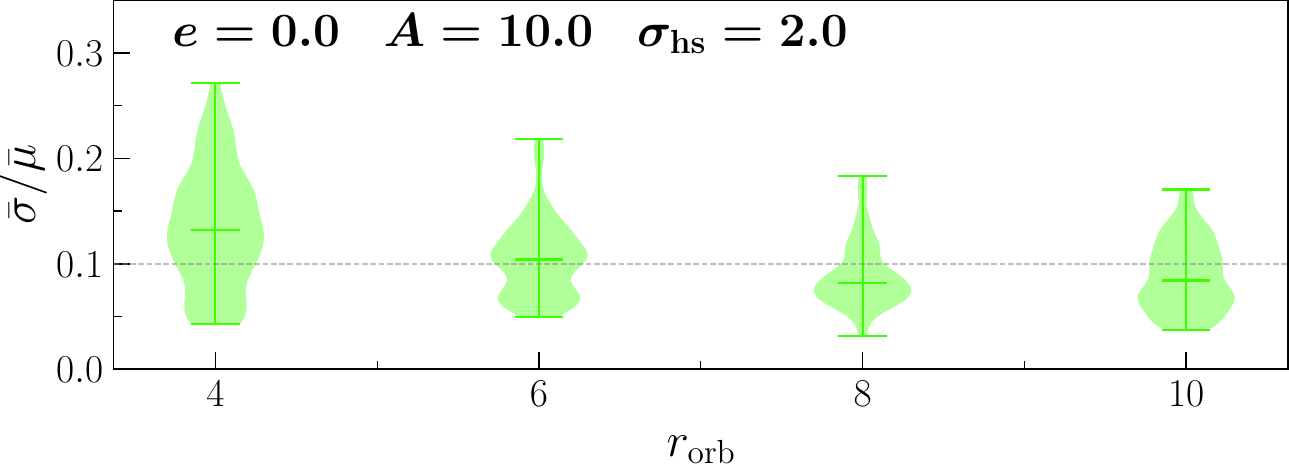}               \begin{picture}(5,5) \put(0,35){\Large (b)}\end{picture}\\
       	\includegraphics[width=0.339\textwidth]{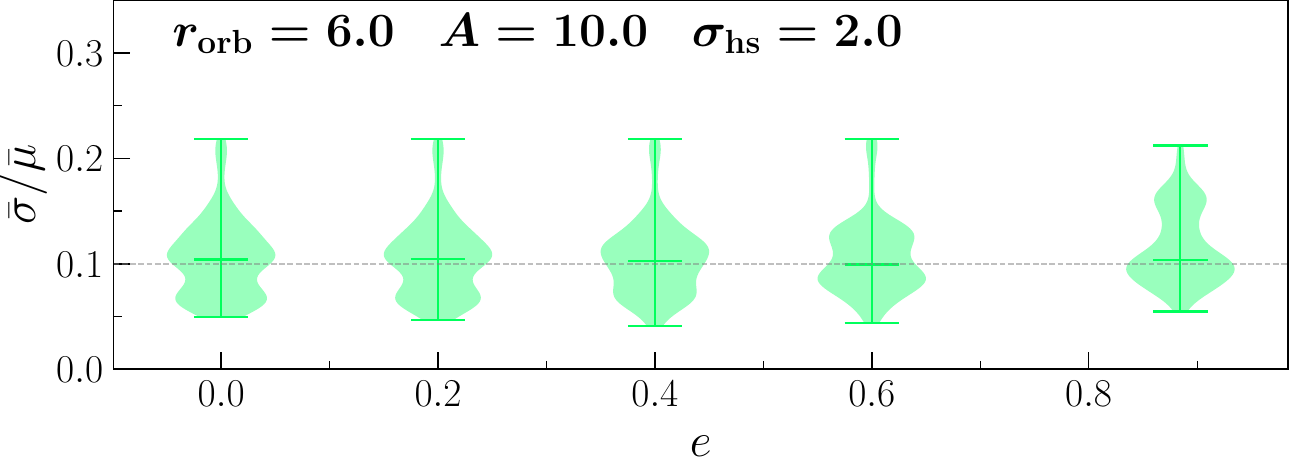}                \begin{picture}(5,5)\put(0,35){\Large (c)}\end{picture}\\
    	\includegraphics[width=0.339\textwidth]{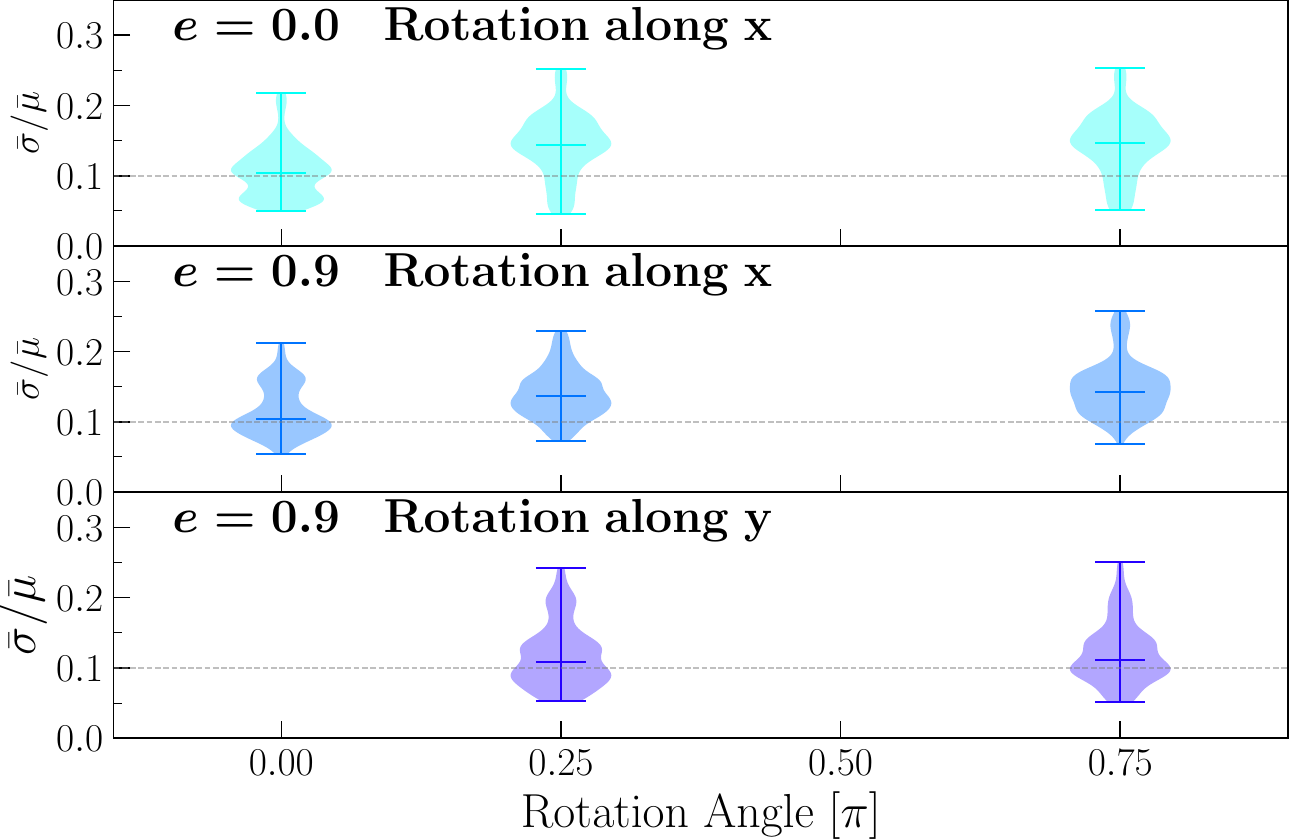}          \begin{picture}(5,5)\put(0,60){\Large (d)}\end{picture}\\
       	\includegraphics[width=0.339\textwidth]{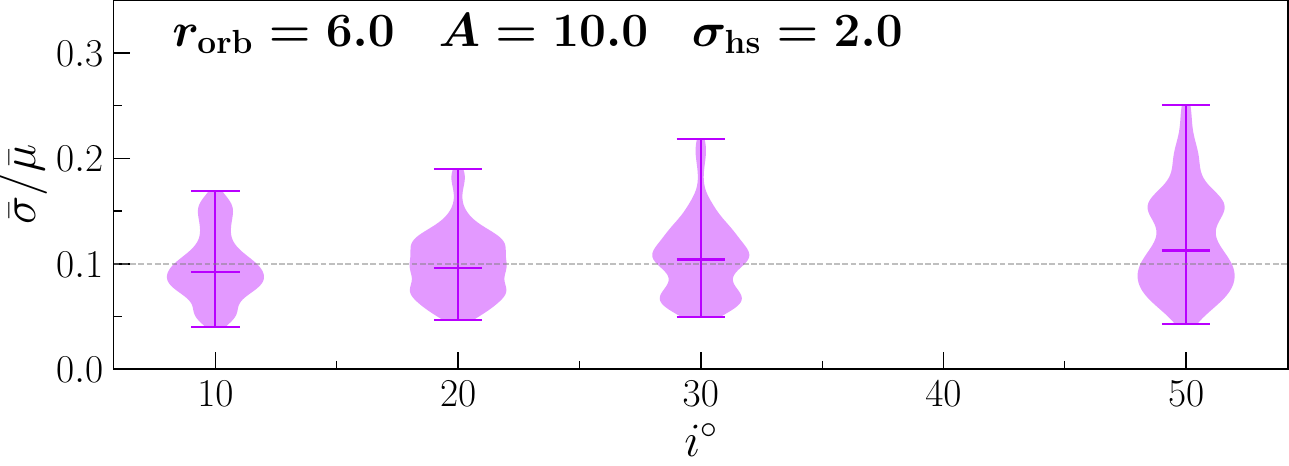}                  \begin{picture}(5,5)\put(0,35){\Large (e)}\end{picture}\\
    	\includegraphics[width=0.339\textwidth]{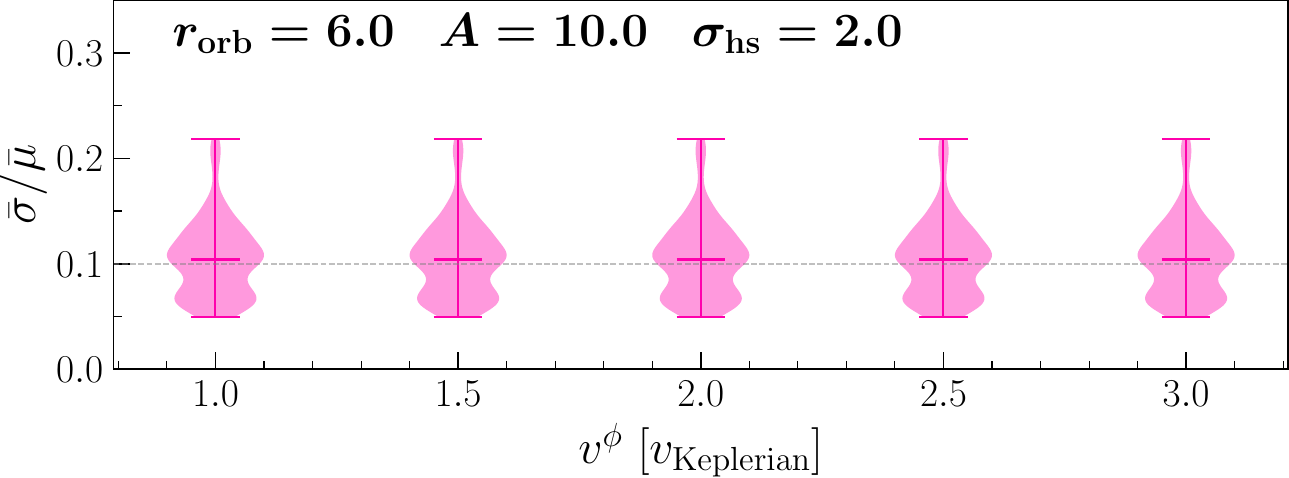}    \begin{picture}(5,5)\put(0,35){\Large (f)}\end{picture}\\
    	\includegraphics[width=0.339\textwidth]{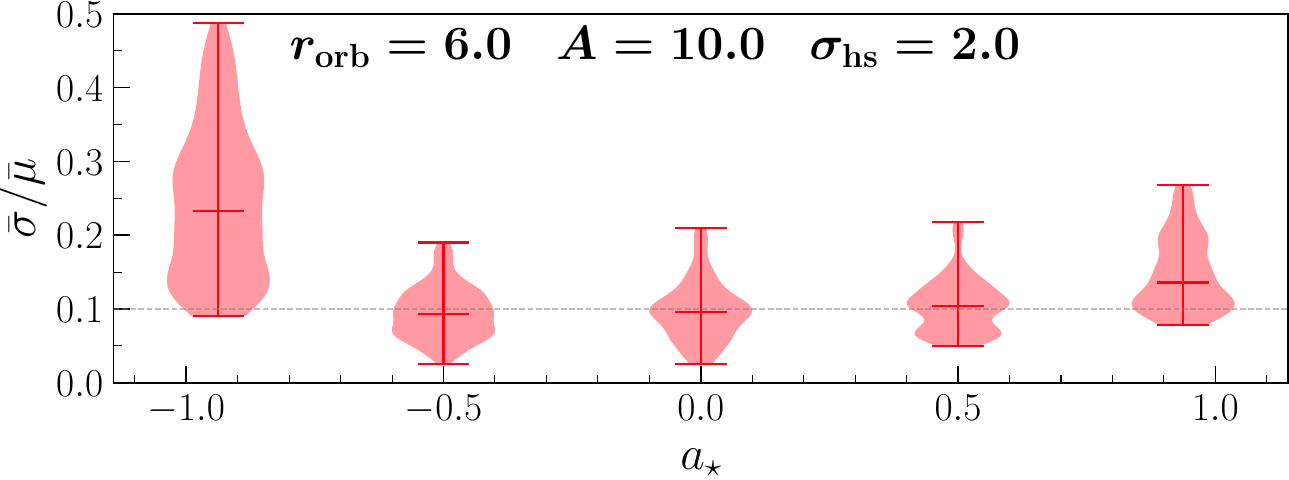}                \begin{picture}(5,5)\put(0,35){\Large (g)}\end{picture}
     \caption{Violin diagrams of the modulation index $\bar{\sigma}/\bar{\mu}$ 
     computed using piecewise sections of 3 hours of the light curve at 
     $230\, {\rm GHz}$. a) Shown the modulation indices varying the hotspot size 
     $\sigma_{\rm hs}$ and its emissivity amplitude $A$, b) changing the orbital 
     radius, c) the orbit eccentricity $e$, d) the rotation angle of the orbital plane, 
     e) the observer view angle $i^{\circ}$,  f) orbital velocity and g) black hole spin, 
     $a_{\star}$, respectively. 
         \label{fig:Violin}}
    \end{figure}

    Recent studies in ideal and resistive GRMHD simulations demonstrate the self-consistent 
    formation of plasmoids within the  funnel accretion flow onto the black hole, as well as 
    within the jet funnel. These simulations reveal the development of current sheets and 
    a highly turbulent plasma morphology, capturing key features of magnetic reconnection 
    and jet dynamics \cite{Nathanail2020,Ripperda2020,Ripperda2022,Vos2024b,Jiang2024,Grehan2025}.  
    
    Motivated by these findings, we introduce a toy model that represents 
    a zeroth-order approximation of a hotspot under the following assumptions. 
    The hotspot does not interact self-consistently with the surrounding plasma 
    and preserves its morphology throughout its orbit. The magnetic-field 
    within the hotspot is assumed to be identical to that of the background plasma. 
    The hotspot follows a stable Keplerian orbit, and its electron distribution function 
    is described by a non-thermal kappa distribution, consistent with that adopted 
    for the plasma. The emission and absorption coefficients within the hotspot 
    are modeled as enhanced versions of those of the background plasma 
    obtained from the GRMHD simulations. This enhancement is prescribed by a 
    three-dimensional Gaussian profile centered at $(x_{\rm orb},\, y_{\rm orb},\, 
    z_{\rm orb})$, as follows:  

    \begin{eqnarray}
    \alpha_{\rm hs} = \alpha\, A\, exp\left[-\frac{r_{\rm hs}^2}{2\sigma_{\rm hs}^{2}}\right],\quad 
    j_{\rm hs} = j\, A\, exp\left[-\frac{r_{\rm hs}^2}{2\sigma_{\rm hs}^{2}}\right],\label{eq:plasmoid}
    \end{eqnarray}
    
    where $r_{\rm hs}^2=(x-x_{\rm orb})^2 + (y-y_{\rm orb})^2+(z-z_{\rm orb})^2$, and 
    $\sigma_{\rm hs}$ is the width of the hotspot, expressed in units of gravitational radii. 
    Here $A$ is the intensity amplitude of the hotspot, which determines the enhancement of the local emission;
     \ie the same profile is used for both the emissivity, and absorptivity coefficients, $j$ and 
     $\alpha$ \cite[for more details see][]{Pandya2016}. The angular velocity of the hotspot is 
     given as function of the Keplerian angular 
     velocity, $\Omega = M^{1/2}/( r_{\rm orb}^{3/2} + a_{\star}M^{1/2})$, and its position as a 
     function of time is described by $(x_{\rm orb},\, y_{\rm orb},\, z_{\rm orb}) = (r_{\rm orb}
     \cos[\Omega\,t],\, r_{\rm orb}\, \sqrt{1 -e^2}\sin[\Omega\, t],\,0)$.
    To take into account for orbital inclination and orientation, we apply a rotation using the 
    matrix $M^{ij}=M^{ij}(\theta,\, \phi,\, \psi)$, where $(\theta,\,\phi,\,\psi)$ are the Euler angles 
    corresponding to the rotations around $x,\, y$ and $z$ axis, repectively.\\ 
    The figure \ref{fig:orbit} shows schematic orbits with different inclination of orbital plane 
    around the black hole at the galactic center, Sgr\,A*. The red arrow indicate the direction 
    of the black hole spin, and $\phi$ is the Euler angle.
    For simplicity, we assume that the hotspot moves in a stable Keplerian orbit, always 
    corotating with the accretion disk, and that its morphology remains unchanged during the 
    evolution. However, its intensity varies in response to changes in the background emission, 
    following the relation in equation \eqref{eq:plasmoid}.
    Our hotspot toy model has seven degrees of freedom. In addition, we consider the black hole 
    spin and the observer’s viewing angle, resulting in nine parameters explored across different 
    scenarios, a total of 37 models: a) Amplitude versus hotspot size, $A=\{1,\,5,\,10,\,15\} $ 
    and $\sigma_{\rm hs}=\{1,\,2,\,3\} \ M$. b) four values of orbital radii 
    $r_{\rm orb}=\{4,\, 6,\, 8,\,10\} \,M$. c) Five eccentricity values $e=\{0,\, 0.2,\, 0.4,\, 0.6,\, 0.9 \}$. 
    d) Three angles $\phi=0,\,\pi/4,\, 3\pi/4$ versus $e=[0,\,0.9]$ and $\theta=0,\,\pi/4,\, 3\pi/4$ 
    for $e=0.9$. e) Four observer inclination angles $i^{\circ}=\{10,\, 20,\, 30,\,50\}$. 
    f) Five hotspot velocities $v^{\phi}=\{1,\,1.5,\,2,\,2.5,\,3\} \, v_{\rm Keplerian}$. 
    g) Five black hole spins $a_{\star}=\{ \pm 15/16,\, \pm 1/2,\, 0\}$.\\
     We choose the following parameters as {\it fiducial} model: 
    $A=10,\, \sigma_{\rm hs}=2,\, r_{\rm orb}=6,\, e=0,\, \phi=0,\, \theta=0,\, i^{\circ}=30,\, 
    a_{\star}=15/16$ and $v^{\phi}=v_{\rm Keplerian}$.

    \section{Constraining the synthetic images} 
    \label{sec:results}
    
    We constrain our models using the main morphological feature of Sgr\,A*, the three hotspots located in 
    the photon ring region. Additionally, we analyse the light-curve variability using the modulation index, 
    $\bar{\sigma}/\bar{\mu}$, where $\bar{\mu}$ is the mean total intensity over a time span of approximately 
    30 hours, and $\bar{\sigma}$ is the corresponding standard deviation. We inspect the synthetic image 
    {\it morphology} across 500 snapshots per model, 13500 images in total, specifically searching for the 
    presence of three distinct {\it hotspots} and lowest variability of $\bar{\sigma}/\bar{\mu} \lesssim 0.1$. 
    
    In Figure \ref{fig:rep}, we show representative snapshots for five different black hole spin values, spanning 
    from counter-rotating to co-rotating cases, by selecting the best approximation to the observed image of 
    Sgr\,A*. Since the image reported by the EHT collaboration represents an average obtained by clustering 
    multiple possible reconstruction algorithms---and therefore includes observational uncertainties and 
    non-physical artifacts---we also present the corresponding time-averaged images, computed over a 
    three-hour window centered on the snapshot shown in the first row.
    
    The top panels show the GRRT images, while the bottom panels display the synthetic images 
    convolved with the EHT resolution, using a beam size of $20\,\mu\text{as}$. Since the hotspot uses the 
    background emission, variations in the spin affect the plasma dynamics and, consequently, the intensity 
    of the hotspot. In the GRRT images, we observe lensing effects due to the presence of the hotspot—such 
    lensing may explain the appearance of the three brightness regions. The time-averaged images (third 
    and fourth rows of Figure \ref{fig:rep}) exhibit a broader emission region surrounding the ring structure 
    which comes from accretion disk averaged dynamics. In the convolved images, only two prominent 
    hotspots become apparent, highlighting the impact of time-averaging and instrumental resolution on 
    the observed morphology.
    
    The variability of the {\it light curve} at 230\,GHz is quantified using the modulation index, as shown 
    in Figure~\ref{fig:Violin}. We present the modulation index distributions using violin plots, which effectively 
    illustrate both the median values and the spread of variability across the different model realizations. The 
    violin plot represents the distribution of the ratio $\bar{\sigma}/\bar{\mu}$, 
    computed over 3-hour intervals while scanning the full 30 hour light-curve for each model. Below, we 
    summarize the main characteristics of each survey:
    
    {\it a) Hotspot size and intensity amplitude.} For hotspots with a size of $\sigma_{\rm hs} = 1\,M$, the 
    modulation index remains below the variability threshold $\bar{\sigma}/\bar{\mu} \lesssim 0.1$ for all 
    tested intensity amplitudes. In the case of $\sigma_{\rm hs} = 2\,M$, the modulation index increases 
    with the amplitude $A$, setting a lower limit of $A \gtrsim 10$ to exceed the threshold. For larger 
    hotspots with $\sigma_{\rm hs} = 3\,M$, amplitudes as low as $A \gtrsim 5$ already produce significant 
    variability. These simulations indicate that larger hotspot with higher intensity contribute to more 
    variability in the plasma emission (see Figure~\ref{fig:Violin}\,a). Moreover, due to the inclination of the 
    viewing angle in the fiducial model ($i=30^\circ$), the black hole shadow appears offset in the image. 
    In this configuration, gravitational lensing becomes more prominent where the photon ring is closer to 
    the shadow from our perspective, and increasing the hotspot size can significantly affect the observed 
    morphology, potentially bringing parts of the emission close to the photon ring.
    
    {\it b) Orbital radius.} The mean modulation index decreases with increasing orbital radius, setting a 
    lower limit for the hotspot location at $r_{\rm orb} \gtrsim 6\,M$ to satisfy the variability constraint 
    $\bar{\sigma}/\bar{\mu} \lesssim 0.1$. This result has two important implications. First, a hotspot located 
    closer to the black hole moves with a higher angular velocity, since we assume a Keplerian velocity 
    profile that decreases with radius. Second, the hotspot must be formed in regions of lower density 
    and temperature, which are typically found farther from the event horizon (see Figure~\ref{fig:Violin}\,b). 
    In the shadow morphology, we observe that when a hotspot is  located closer to the black hole, the 
    relativistic effects are stronger, the lensing effects and Doppler beaming enhances its brightness 
    asymmetrically.
    
    {\it c) Orbital eccentricity.} In Figure~\ref{fig:Violin}\,c, the mean value 
    of the modulation index across different eccentricities remains around $\bar{\sigma}/\bar{\mu} \sim 0.1$. 
    The distribution appears bimodal, with a preference for higher eccentricities. This behaviour may 
    arise because, in highly eccentric orbits, the hotspot spends most of its time at larger distances from 
    the black hole, where relativistic effects such as gravitational lensing and Doppler beaming are weaker, 
    leading to lower variability. However, as the hotspot approaches pericenter, it experiences rapid 
    acceleration and stronger relativistic effects, resulting in a brief but enhanced variability in the light curve.
    Moreover, in highly eccentric orbits, the hotspot is moving from the  perihelion and the aphelion or the 
    orbit, leading to a more asymmetric and variable structure. In contrast, for circular orbits (zero eccentricity), 
    the hotspot remains at same radii, producing a more symmetric image.
     
     {\it d) Inclination of the orbital plane.} The fiducial model assumes that the hotspot orbit lies in the 
     equatorial plane of the black hole. To explore the impact of inclination, we apply the rotation matrix 
     $M^{ij}(\phi,\, \theta,\, \psi)$ to the fiducial orbit, choosing angles $\theta = \pi/4$ and $3\pi/4$ (first 
     and second rows of Figure~\ref{fig:Violin}\,d), consistent with the setup used in \cite{Yfantis2024}, 
     and $\phi = \pi/4$ and $3\pi/4$ (third row of Figure~\ref{fig:Violin}\,d). The results indicate a clear 
     preference for $\theta = 0$, regardless of eccentricity, suggesting that orbits closer to the equatorial 
     plane produce less variability. In contrast, for $\phi$ rotations at high eccentricity ($e = 0.9$), we 
     find a mean modulation index of $\bar{\sigma}/\bar{\mu} \sim 0.1$, indicating moderate variability 
     under these configurations.
     Moreover, for large orbital inclination angles and high eccentricities, the hotspot periodically appears 
     and disappears from the shadow image, resulting in a more asymmetric and variable structure. In 
     contrast, for orbits confined to the black hole equatorial plane, the hotspot remains continuously within 
     the observer field of view, producing a more symmetric and stable image. The most pronounced effect 
     is observed for the model with $\theta=0$, $\phi=3\pi/4$, and $e=0.9$, where the hotspot remains 
     hidden for most of the time.
     
     {\it e) Viewing angle of the observer.} The best-fit viewing angle from EHT observations is $i^{\circ} = 30$ 
    \cite{EHT_SgrA_PaperV}, which we adopt as our fiducial model. However, we also explore three 
    additional inclination angles, given that the observed black hole shadow is sensitive to the observer 
    line of sight due to Doppler beaming and gravitational lensing effects. 
    Changes in the viewing angle alter the projection of the hotspot orbit and, due to gravitational lensing, also 
     affect the apparent shape of the hotspot. When the hotspot passes through regions farther from the black 
     hole, it appears more elongated. In Figure~\ref{fig:Violin}\,e, we present the distribution of the modulation 
     index for the different viewing angles. Based on the variability threshold, our results suggest a preference 
     for $i^{\circ} \lesssim 30$, consistent with EHT results, while higher inclinations, such as $i^{\circ} = 50$, 
     exhibit increased variability.

    \begin{figure*}
    \centering
    	\hspace{1.25cm}\includegraphics[width=0.80\textwidth]{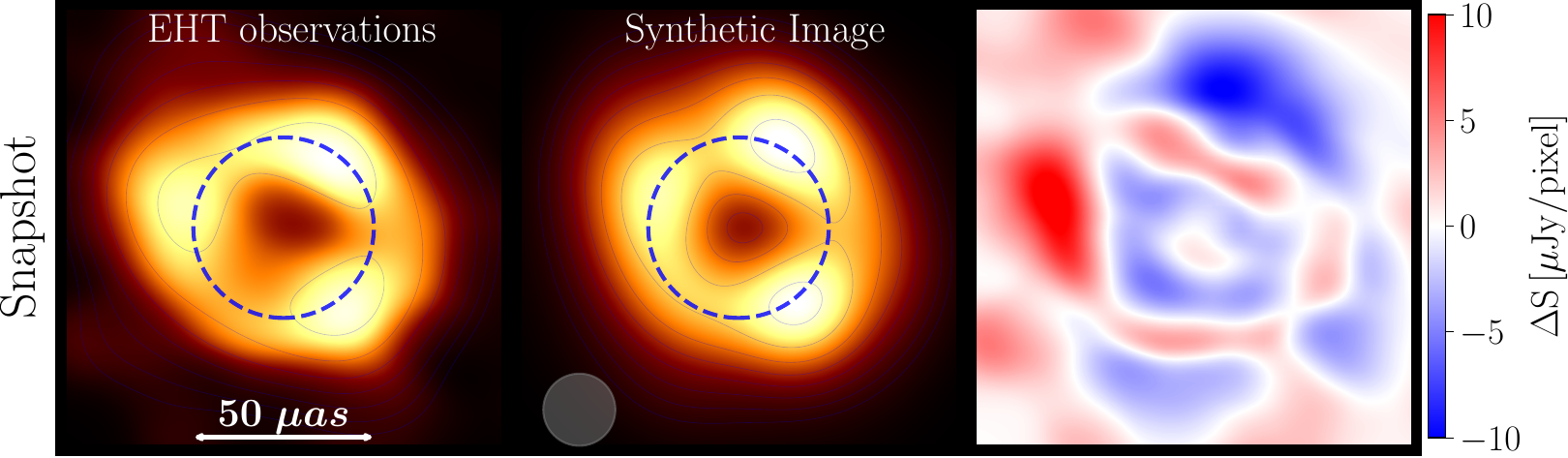}\\
    	\hspace{1.25cm}\includegraphics[width=0.80\textwidth]{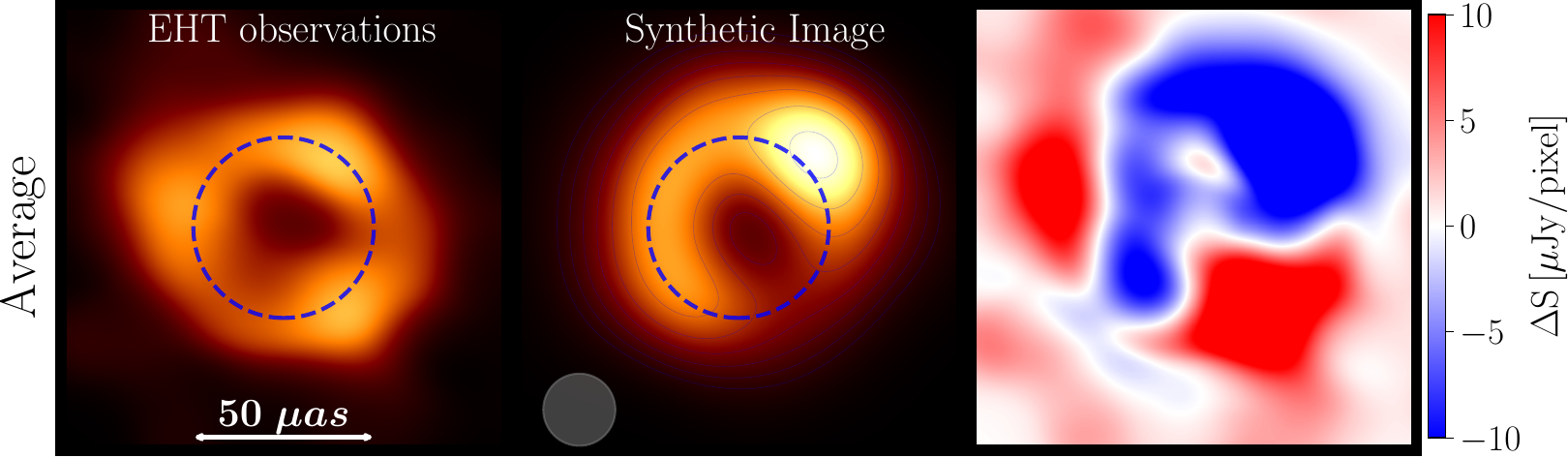}\\
    	\includegraphics[width=0.738\textwidth]{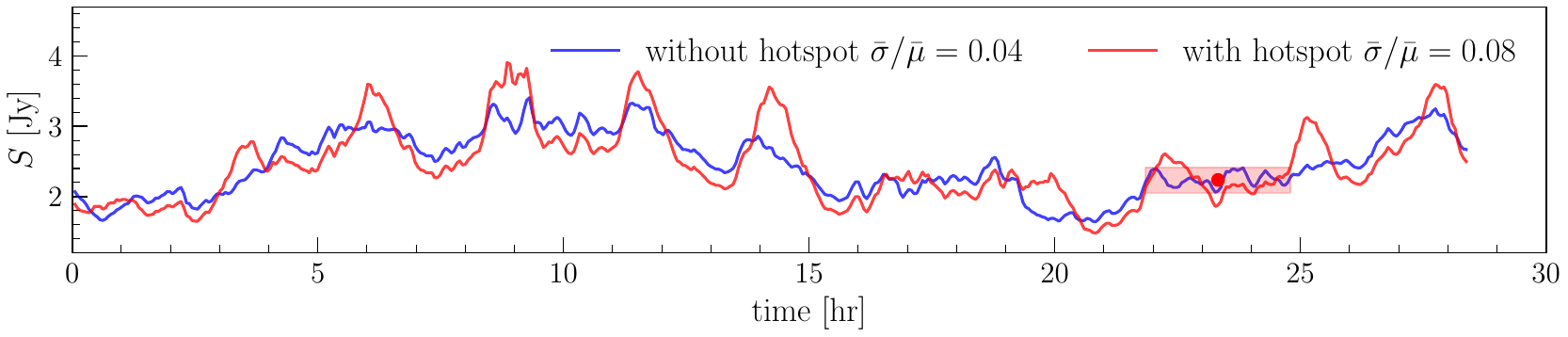}
     \caption{Comparison between the observed image of Sagittarius A* by the 
     EHT collaboration at $230\,{\rm GHz}$ and synthetic images from GRMHD 
     and GRRT simulations. The first row shows a snapshot corresponding to 
     the red dot in the light curve, while the second row presents the time-averaged 
     image computed over a three-hour window indicated by the red shaded region. 
     The third row displays the light curve of the total flux density; the blue curve 
     represents the emission evolution without hotspot, while the red curve 
     includes the contribution from orbiting hotspot, the modulation index, 
     $\bar{\sigma}/ \bar{\mu} < 0.1$ in both cases. We depict the relative error 
     between observation and synthetic images $\Delta S= I_{\rm obs}-I_{\rm syn}$ 
     in the third column. This model corresponds to a rotating black hole with dimensionless 
     spin $a_{\star}=0.5$ observed by a camera at  inclination view angle $i^{\circ}=20$.
    \label{fig:best}}
    \end{figure*}

    {\it f) Hotspot angular velocity.} As discussed in Section~\ref{sec:hotspot}, our fiducial model assumes 
    that the hotspot moves with Keplerian angular velocity. In Figure~\ref{fig:Violin}\,f, we present the 
    modulation index for five models with different angular velocities $v^{\phi}$, expressed in terms of the 
    Keplerian velocity $v_{\rm Keplerian}$. The results show a mean value of $\bar{\sigma}/\bar{\mu} \sim 0.1$ 
    with a consistent distribution across models. This behavior may be attributed to the hotspot proximity 
    to the event horizon, where strong magnetic fields are typically present in the accretion flow. However, 
    the hotspot appearance remains unchanged regardless of the angular velocity, and the resulting image 
    closely resembles that of the fiducial model morphology.
    
    {\it g) Black hole spin.}  We assume that the direction of hotspot motion is consistent across all models, 
    co-rotating with the accretion disk and may exhibit either co-rotating or counter-rotating 
    motion relative to the black hole spin. 
    For negative spin values, a counter-rotating hotspot experiences frame-dragging that acts against its 
    motion, resulting in an elongation of its orbital path. In contrast, a co-rotating hotspot—associated with a 
    positive spin—benefits from frame-dragging, which reduces its orbital period. As shown in Figure~\ref{fig:rep}, 
    this effect enhances the emission, producing the brightest hotspot in the co-rotating case. These relativistic 
    effects become more pronounced at higher spin magnitudes, $a_{\star} = \pm 0.94$, leading to a significant 
    increase in variability, as illustrated in Figure~\ref{fig:Violin}\,g.
    The modulation index shows enhanced variability for counter-rotating hotspots, particularly in the case 
    $a_{\star} = -0.94$, where all values exceed the variability threshold due to the stronger changes in 
    brightness. This suggests that black holes with high spin magnitudes are associated with greater variability, 
    implying a lower bound on the spin parameter of $|a_{\star}| \gtrsim 0.5$. In contrast, for intermediate and 
    zero spin cases, the modulation index generally decreases, reflecting weaker frame-dragging effects and 
    reduced asymmetries in the emission, which translates into lower variability.
    
    {\it Preferred model.} In Figure~\ref{fig:best}, we present a snapshot of the synthetic image (marked 
    by a red dot in the light curve) that provides the most likely model to the EHT observations of the supermassive 
    black hole at the Galactic Center. The second row shows the corresponding time-averaged image, 
    computed over a three-hour window indicated by the red shaded region in the light curve. The model 
    assumes a dimensionless spin parameter of $a_{\star}=0.5$, a magnetically arrested disk (MAD) 
    configuration for the magnetic field, and a viewing inclination angle of $i^{\circ}=20$. The hotspot emission 
    is embedded within the background synchrotron nonthermal radiation with intensity amplitude of $A=10$, 
    where we assume that fifty percent of the magnetic energy ($\epsilon=0.5$) is used to heat the electrons 
    in the plasma. The hotspot has a horizon-scale size, \ie $\sigma_{\rm hs}=2\,M$. Our results indicate that 
    the hotspot is moving on a circular orbit ($e=0$) at a radius of $r_{\rm orb}=6\,M$, with a Keplerian angular 
    velocity, lying in the equatorial plane of the black hole ($\phi,\, \theta,\, \psi=0$). 
    In the top panels of Figure~\ref{fig:best}, we show the observed image of Sgr A$^*$ obtained by the EHT 
    collaboration in 2017, the best-fit synthetic image from our theoretical model convolved with the same 
    observational resolution ($20\,\mu{\rm as}$), and the pixel-by-pixel relative difference between them. 
    The dotted blue circle indicates the approximate location of the photon ring at $\sim 50\,\mu{\rm as}$. 
    The synthetic image successfully reproduces the characteristic three hotspots near the photon ring, with 
    differences on the order of $10\,\mu{\rm Jy}$. Additionally, we computed several metrics ({\tt MSE}, {\tt SSIM}, 
    {\tt DSSIM}, and {\tt NCC}) to quantify the similarity between the observed and synthetic images, selecting 
    the best model based on the minimum mean squared error, {\tt MSE} $\sim 0.0037$. 
    For completeness, we computed the time-averaged image over a three-hour observation window. In this 
    case, the image shows a dominant hotspot within the photon ring and a nearly continuous ring structure 
    with a small intensity depression. Comparing our results with this averaged scenario is more appropriate, 
    given that the EHT-reconstructed image represents a time-averaged observation rather than a single 
    snapshot. Despite the introduction of the orbiting hotspots, our model is still unable to 
    reproduce the three distinct hotspots observed in the EHT image. In the bottom panel, we present the light 
    curve of the best-fit model (in red) together with the corresponding model without a hotspot (in blue). For 
    the preferred model, the global modulation index is $\bar{\sigma}/\bar{\mu} \sim 0.1$, however, focusing locally 
    within a 3 hour window around the snapshot (marked with a red dot), we find $\bar{\sigma}/\bar{\mu} = 0.08$ 
    and $\bar{\sigma}/\bar{\mu} = 0.04$ for both the model with and without the hotspot, respectively.
    
    \section{Summary and Conclusions} \label{sec:conclusions}
    
    In this work, we systematically studied the image of the supermassive black hole Sgr\,A$^*$, focusing on the 
    presence of hotspots near the photon ring, as observed by the Event Horizon Telescope Collaboration 
    \cite{EHT_SgrA_PaperI}, and the variability of the 230\,GHz light-curve, characterized by the modulation 
    index $\bar{\sigma}/\bar{\mu}$. Since state-of-the-art GRMHD and GRRT simulations are unable to 
    reproduce these features, we propose the presence of an orbiting hotspot around the black hole. Assuming 
    nonthermal synchrotron emission, we explored a variety of scenarios to investigate the effects of hotspot 
    size and intensity, angular velocity, orbital radius, eccentricity, orbital plane inclination, black hole spin, 
    and observer viewing angle on the resulting image and variability.
    
    We analyzed 13,500 synthetic images, searching for configurations that reproduce the characteristic three 
    hotspots in the black hole shadow morphology and evaluating the corresponding light-curve variability. We 
    found that gravitational lensing is strong in all cases and may naturally produce multiple bright regions by 
    lensing a hotspot generated from turbulent plasma in the disk, winds and accretion funnel. From our 
    parameter survey, several important trends emerged. In terms of hotspot size and intensity, larger and 
    brighter hotspots tend to increase the modulation index, with intensity amplitudes constrained to 
    $A \lesssim 10$ to remain consistent with observational variability limits. Regarding the orbit, we 
    found that hotspots located within $r_{\rm orb} \lesssim 6\,M$ exhibit higher modulation indices. Although 
    orbital eccentricity has a limited effect on the overall variability, circular orbits provide the best match to 
    the observed morphology, particularly when the hotspot moves along the black hole equatorial plane at 
    approximately Keplerian velocities.
    
    In addition, we found that plasmas surrounding highly spinning black holes ($|a_{\star}| \gtrsim 0.5$) tend 
    to be more turbulent, leading to stronger light-curve variability. Finally, the observer viewing inclination plays 
    a significant role: higher inclination angles enhance both the variability and the distortion of the hotspot 
    shape due to stronger gravitational lensing effects.
    
    After applying the two observational constraints---the presence of three hotspots in the black hole image 
    and the modulation index of the light curve---to snapshots, we found that the preferred model corresponds 
    to a black hole with dimensionless spin $a_{\star}=0.5$, a MAD (magnetically arrested disk) configuration 
    for the magnetic field, and a hotspot of size $\sigma_{\rm hs}=2M$ with an intensity amplitude $A=10$. 
    The hotspot moves in a circular orbit at $r_{\rm orb}=6M$ with Keplerian velocity in the black hole equatorial 
    plane and is viewed by the observer located at the earth at an inclination of $i^\circ=20$, which is slightly 
    different from the value reported by the EHT Collaboration \cite{EHT_SgrA_PaperV}. This model successfully 
    reproduces the three characteristic hotspots observed by the EHT, achieves a low mean squared error of 
    ${\tt MSE} \sim 0.0037$ between the synthetic and observed images, and satisfies the variability constraint 
    with a global modulation index of $\bar{\sigma}/\bar{\mu} \lesssim 0.1$. Quantitative comparison metrics, 
    such as the structural similarity index (SSIM) and the normalized cross-correlation (NCC), further confirm 
    the close agreement between the synthetic and observed images. 
    
    To avoid overinterpreting our numerical results, we compute the time-averaged image of the preferred 
    model covering a three hour observation window to match the EHT imaging procedure for Sgr\,A*. Despite 
    the presence of multiple transient structures in individual snapshots, the time-averaged image reveals only 
    a single dominant hotspot. This result suggests that, 
    although the snapshot successfully reproduces the three observed hotspots, time-averaging tends to smooth 
    out transient features, reducing the number of prominent hotspots in the averaged synthetic image.
    These results provide constraints on the geometry and dynamics of the emitting plasma around Sgr\,A*, 
    supporting the interpretation of the observed structure as driven by gravitational lensing, orbital motion, 
    and plasma dynamics, including hotspot formation and microscopic plasma processes near the event 
    horizon, also highlight the role of magnetic reconnection in driving electron heating and acceleration. 
   
    Nevertheless, our results are not conclusive, and our toy model has some caveats. In particular, 
    the stable hotspot is introduced ad hoc rather than arising self-consistently from the plasma dynamics 
    in our GRMHD simulations.
    Furthermore, we considered only a limited set of black hole spin values. Additionally, the three 
    hotspots inferred from EHT black hole image may be artifacts of the combined imaging methods.
    Future studies should investigate hotspot formation and evolution using resistive GRMHD or GR-PIC 
    simulations. In addition, a more comprehensive analysis incorporating slow-light ray-tracing, polarimetry, 
    and astrometry \cite{Wielgus2022,Vos2022,Antonopoulou2024,Yfantis2024} will be required to robustly 
    interpret the presence of hotspots in the upcoming VLBI observations.

\section*{Acknowledgments}
    This research was supported by DGAPA-UNAM (grant IN110522) and the Ciencia 
    Básica y de Frontera 2023–2024 program of SECIHTI México (projects 
    CBF2023-2024-1102, 257435, 1147615 and 1080312). 
    Simulations were performed at Laboratorio Nacional de Supercómputo del Sureste 
    de México (projects 202304071C  and 202403062N), at Atocatl cluster of LAMOD-DGTIC 
    UNAM, ‘‘Going Merry'' and "Thousand Sunny" workstations.
    
%------------------------------------------------------------------------
\section*{Data Availability}
%------------------------------------------------------------------------
The data underlying this article will be shared on reasonable request
to the corresponding author.

%\bibliographystyle{mnras}
%\bibliography{aeireferences} % if your bibtex file is called example.bib

%\label{lastpage}
\end{document}